\documentclass[balance]{article}
\usepackage[T1]{fontenc} 
\usepackage[utf8]{inputenc} 
\usepackage{ismir,amsmath,cite,url}
\usepackage{graphicx}
\usepackage{color}
\def\lbd{}	
\usepackage{units}
\usepackage{amssymb}
\usepackage{comment}
\usepackage{booktabs}
\usepackage{siunitx}
\usepackage{lineno}
\usepackage{tabularx}

\usepackage{pifont}
\renewcommand{\checkmark}{\ding{51}}
\newcommand{\xmark}{\ding{55}}%
\title{AVASpeech-SMAD: A Strongly Labelled Speech and Music Activity Detection Dataset with Label Co-Occurrence}

\newcommand*{\authsep}{\hspace{1cm}}
\multauthor
{
    Yun-Ning Hung$^1$ \authsep
    Karn N. Watcharasupat$^{1, 2}$ \authsep
    Chih-Wei Wu$^3$ \authsep
    Iroro Orife$^3$
}{\bfseries{
    Kelian Li$^1$ \authsep
    Pavan Seshadri$^1$ \authsep
    Junyoung Lee$^2$
}\\
    $^1$Center for Music Technology, Georgia Institute of Technology, USA\\
    $^2$School of Electrical and Electronic Engineering, Nanyang Technological University, Singapore\\
    $^3$Netflix, Inc., USA\\
    {Corresponding author: {\tt amy.hung@gatech.edu} (Y.-N. Hung)}
}

\def\authorname{Y.-N. Hung, K. N. Watcharasupat, C.-W. Wu, I. Orife, K. Li, P. Seshadri, and J. Lee}

\begin{document}

\maketitle
\begin{abstract}
We propose a dataset, AVASpeech-SMAD, to assist speech and music activity detection research. With frame-level music labels, the proposed dataset extends the existing AVASpeech dataset, which originally consists of \num{45} hours of audio and speech activity labels. To the best of our knowledge, the proposed AVASpeech-SMAD is the first open-source dataset that features strong \emph{polyphonic} labels for both music and speech.
The dataset was manually annotated and verified via an iterative cross-checking process. A simple automatic examination was also implemented to further improve the quality of the labels. Evaluation results from two state-of-the-art SMAD systems are also provided as a benchmark for future reference.   
\end{abstract}
\section{Introduction}\label{sec:introduction}

Speech and music activity detection (SMAD) is a long-studied problem
and has been included in the Music Information Retrieval Evaluation eXchange (MIREX) competition for several years. 
Like other Music Information Retrieval (MIR) tasks, the most recent improvements to SMAD rely on data-driven approaches 
\cite{venkatesh2021investigating, lemaire2019temporal, de2019exploring, jang2019music, ddoukhanmirex2018, papakostas2018speech}. However, due to copyright issues, many of these systems were trained on private datasets which impeded the reproducibility of the results (e.g., the radio datasets used in \cite{lemaire2019temporal} and \cite{venkatesh2021investigating}). Although several publicly available datasets have been proposed for training or evaluation of SMAD systems, as shown in Table~\ref{tab:dataset}, they suffer from several drawbacks. GTZAN \cite{tzanetakis2000marsyas}, MUSAN \cite{snyder2015musan}, SSMSC \cite{scheirer1997construction} and Muspeak \cite{muspeak} only have non-overlapping speech or music segments. OpenBMAT \cite{melendez2019open} and ORF TV \cite{seyerlehner2007automatic} only support music labels, while the original AVASpeech dataset \cite{chaudhuri2018ava} only has speech labels. 
In the real-world where speech and music co-occur regularly, these datasets might not be suitable training sources.

To solve the data limitation problem, we proposed a supplementary dataset, \textit{AVASpeech-SMAD}. The dataset is the extension of the AVASpeech dataset proposed by Chaudhuri et al.\@ \cite{chaudhuri2018ava}. 
The original AVASpeech dataset only contains speech activities and we extended it by manually labelling the music activities. We expect our proposed dataset to be used for training or evaluation of future SMAD systems. The dataset is open-sourced on the Github repository: \url{https://github.com/biboamy/AVASpeech_Music_Labels}

\begin{table}[t]
\small
\begin{tabularx}{\columnwidth}{X*{3}{S[detect-mode=false, mode=text, table-format=3.2]}}
\toprule
                & {Avg (\%)} & {Min (\%)} & {Max (\%)} \\
\midrule
    Speech      & 52.48 & 14.40 & 87.02 \\
    Music       & 43.40 & 00.29 & 100.00 \\
    Both speech and music     & 20.73 & 00.00 & 77.19 \\
\bottomrule
\end{tabularx}
\caption{Minimum, maximum, and average percentages of regions with speech only, with music only, and with both speech and music for each audio sample.}
\label{tab: statistic}
\end{table}
\vspace{-10pt}

\begin{table*}[t]
\centering
\setlength{\tabcolsep}{3pt}
\small
\begin{tabularx}{\textwidth}{
    Xccc
    S[detect-mode=false, mode=text, table-format=4]
    r}
    \toprule
        Dataset
        & Music Labels 
        & Speech Labels 
        & Overlap 
        & \multicolumn{1}{r}{\# instances} 
        & \multicolumn{1}{r}{Duration (hrs)}\\
    \midrule
    GTZAN Speech and Music \cite{tzanetakis2000marsyas} &  \checkmark  & \checkmark & No & 128 & 1.1   \\
    MUSAN \cite{snyder2015musan} 
        &  \checkmark  & \checkmark & No & 2016 & 109   \\
    Scheirer \& Slaney Music Speech (SSMSC) \cite{scheirer1997construction} 
        &  \checkmark  & \checkmark & No & 245 & 1   \\
    Muspeak \cite{muspeak}          
        & \checkmark & \checkmark & No & 214 & 5   \\
    OpenBMAT \cite{melendez2019open}         
        &  \checkmark  &  \xmark & Yes & 1647& 27.5   \\
    ORF TV \cite{seyerlehner2007automatic}           
        & \checkmark & \xmark & Yes & 13 & 9  \\
    AVASpeech-SMAD (Proposed)         
        & \checkmark & \checkmark & Yes & 160 & 45   \\
    \bottomrule
\end{tabularx}
\caption{Metadata of the existing datasets compared to our proposed AVASpeech-SMAD dataset.}
\label{tab:dataset}
\end{table*}

\section{Dataset}\label{sec:dataset}
The statistics of the dataset are shown in Table~\ref{tab:dataset}. Compared to other publicly available datasets, ours is the only one containing overlapping speech \textit{and} music frame labels. The dataset includes a variety of content, languages, genres and production quality. The audio data, labels, and annotation process are discussed in the following sections.

\subsection{Audio \& Labels}
The original dataset contains \num{160} excerpts of \num{15}-minute clips taken from YouTube videos with a total duration of \SI{45}{\hour}. Each audio file is stereo and was sampled at \SI{22050}{\hertz} with \SI{16}{bits} per sample. Due to copyright issues, we could not distribute the audio files from this dataset. Instead, we include the scripts for downloading and preprocessing the audio files in our GitHub repository. 

The speech labels were derived as-is from the original AVASpeech while the new music labels were manually annotated. The statistics of the labels are shown in Table~\ref{tab: statistic}. The dataset covers a variety of cases. Some samples barely have any music while some contain mostly music. Some of them do not have any region with overlapping music and speech, while some have over \SI{70}{\percent} of the regions containing both music \textit{and} speech. 
The detailed statistics of each clip can be found in our GitHub repository.

\begin{figure}[t]
  \centering
  \includegraphics[width=0.9\columnwidth]{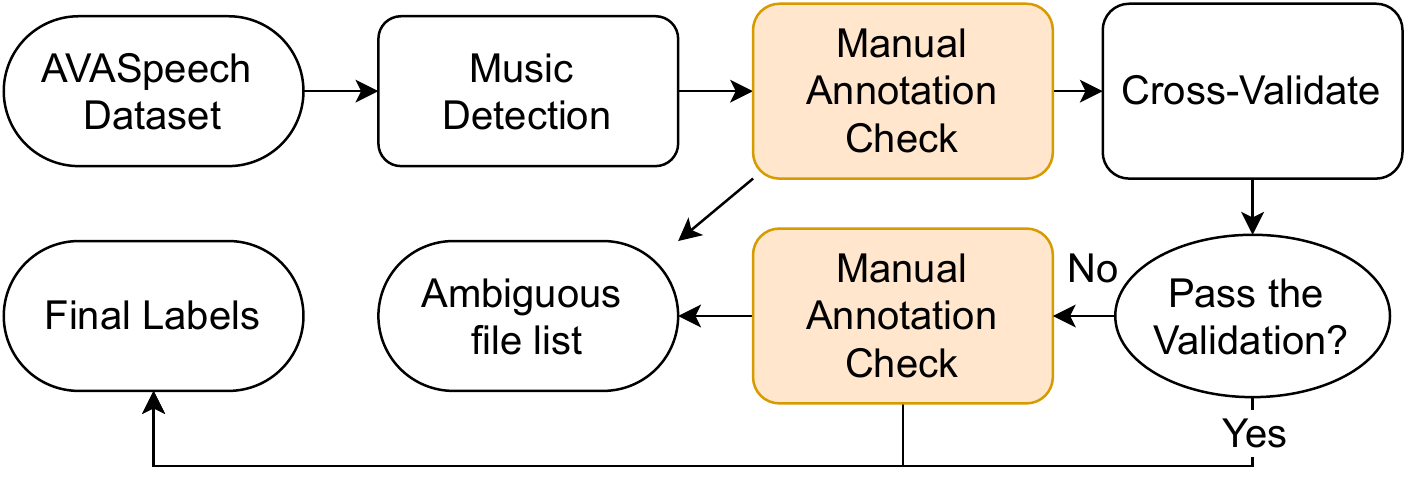}
  \vspace{-10pt}
  \caption{Overview of the annotation process.}
  \label{fig:pipeline}
\end{figure}

\subsection{Annotation Process}
The annotation process of the proposed dataset is shown in Fig.~\ref{fig:pipeline}. The annotation process has four steps, namely music detection, first manual annotation check, cross-validate and second manual annotation check. Steps involving human annotators are highlighted in orange.  Seven MIR students and researchers, all of whom are also musicians of varying experience levels, have volunteered as annotators. An internal algorithm is first used to pseudo-label the music regions. The annotators are then asked to manually annotate the regions with music activity, with the pseudo labels acting as a rough guide. Each annotators were assigned between \num{12} to \num{24} audio clips to annotate active music regions. Sonic Visualizer\footnote{\url{https://www.sonicvisualiser.org}, last accessed 04/30/2021} was used to mark the active regions and all annotators used headphones during the annotation process. To ensure the consistency of the labels and avoid ambiguity, the guideline for determining music versus speech is listed in GitHub repository.


After the annotations are completed, all the labels were cross-validated with the speech labels provided by original AVASpeech dataset. Since the original labels contain ``Speech with music'' and ``Clean speech'' classes, we can expect that the former contains music while the latter do not contains music. Any discrepancies between our labels and the original labels were algorithmically detected: the region that was labeled as music in the original labels but not in our labels, and the region that was labeled no music in the original labels but labeled as music in our labels. If the discrepancies are less than \SI{3}{\second}, we automatically modify the labels based on the original labels. However, if the discrepancies are greater than three seconds, we conducted a manual review of the regions. Each region is randomly assigned to additional two annotators and majority vote is considered to determine the final labels of the regions.  

\begin{table}[t]
\centering
\small
\setlength{\tabcolsep}{3pt}
    \begin{tabularx}{\columnwidth}{X *{6}{S[detect-mode=false,mode=text,table-format=2.1]}}
    \toprule
    & \multicolumn{3}{c}{Music} & \multicolumn{3}{c}{Speech} \\    
    \cmidrule(lr){2-4}\cmidrule(lr){5-7}
    Algo. & {F1} & {Precision} & {Recall} & {F1} & {Precision} & {Recall}\\
    \midrule
    \cite{venkatesh2021investigating}  
        & 80.2 & 69.8 & 94.3 & 77.6 & 82.7 & 73.1 \\
    \cite{ddoukhanmirex2018} 
        & 62.2 & 85.7 & 48.8 & 79.1 & 83.1 & 75.4 \\
    \bottomrule
\end{tabularx}
\caption{The segment-level evaluation (\%) on our AVASpeech-SMAD dataset by using two SOTA systems.}
\label{tab: segment_eval}
\end{table}

\section{Benchmark}
We choose two existing SOTA systems to evaluate on our proposed dataset. First is the CRNN-based detector proposed by \cite{venkatesh2021investigating}. The method is trained on the combination of synthetic data and radio broadcast. The second system is the InaSpeechSegmentor proposed by \cite{ddoukhanmirex2018}. The segmentor has a CNN architecture and can split audio into homogeneous music, speech and noise region. The \texttt{sed\_eval} toolbox is used to perform segment-level evaluation as used in the MIREX 2018 competition\footnote{\url{https://www.music-ir.org/mirex/wiki/2018:Music_and/or_Speech_Detection}, last accessed 04/30/2021}. The result is shown in Table~\ref{tab: segment_eval} for future reference. 

\section{Conclusion}
In this work, we proposed a supplementary dataset for SMAD. The dataset not only contains overlapping speech and music frame labels, but also includes a variety of content. 
Based on our benchmark experiments, one of the SOTA systems achieved F1-scores (80\% and 77 \% for music and speech, respectively) that are slightly lower than the reported 85\% for both music and speech on MIREX test sets in \cite{venkatesh2021investigating}. This result suggests that our proposed dataset is able to present new challenges to the model and potentially complement the existing datasets (e.g., MIREX test sets) with diverse content and frame-level labels. We expect this dataset to serve as a new resource and reference for future SMAD research.



\section{Acknowledgement}
K. N. Watcharasupat and J. Lee acknowledge the support from the CN Yang Scholars Programme, Nanyang Technological University, Singapore.
We also gratefully acknowledge Professor Alexander Lerch and Music Informatics Group supported this research by providing  a  Titan  X  GPU  for computing the experiment.


\bibliography{ISMIRtemplate}
\end{document}